\documentclass{ws-procs9x6}

\usepackage{graphicx}
\usepackage{psfrag}

\def\gsim{\mathrel{\lower3pt\hbox{$\sim$}}\hskip-11.5pt\raise3pt\hbox{$>$}\;}

\def\lsim{\mathrel{\lower3pt\hbox{$\sim$}}\hskip-11.5pt\raise3pt\hbox{$<$}\;}

\begin{document}

\title{A LIGHT SCALAR WIMP, THE HIGGS PORTAL AND DAMA}

\author{MICHEL H.G. TYTGAT}

\address{Service de Physique Th\'eorique\\
Universit\'e Libre de Bruxelles\\
Boulevard du Triomphe, CP225, Brussels 1050, Belgium\\}

\begin{abstract}
In these proceedings, we report on the possible signatures of a
light scalar WIMP, a dark matter candidate with $M_{DM} \sim$ few
GeV, which is supposed to interact with the Standard Model
particles through the Higgs, and which might be related to the
annual modulation observed by DAMA.
\end{abstract}

\bodymatter

\section{Introduction}\label{sec:intro}

There are many models of dark matter but the most acclaimed is the neutralino.
However the most economical extension of the Standard Model with dark matter
consists simply in adding a real singlet scalar field $S$,
\begin{equation}
\label{lag}
{\cal L} \owns \frac{1}{2}\partial^\mu S \partial_\mu S-\frac{1}{2}\mu^2_S \,S^2 -\frac{\lambda_S}{4} S^4 -\lambda_L\, H^\dagger H\, S^2
\end{equation}
where $H=(h^+ \, (h+iG_0)/\sqrt{2})^T$ is the Higgs doublet.
This Lagrangian has a discrete $Z_2$ symmetry, $S \rightarrow - S$ and,
if this symmetry is not spontaneously broken, the $S$ particle
is a dark matter candidate
\cite{McDonald:1993ex,Burgess:2000yq,Barger:2007im},
with mass
\begin{equation}
\label{massS}
m_S^2=\mu^2_S+\lambda_L \mbox{\rm v}^2,
\end{equation}
where $\hbox{v}=246$~GeV. This is also one of the simplest instance
of dark matter through the Higgs portal, a general scheme according
to which a hidden sector interact with ordinary matter through the
Higgs sector of the Standard Model\cite{Patt:2006fw}. Also the
singlet scalar extension effectively encompasses many other models
with extra fields, for instance the so-called Inert Doublet Model
(IDM)
\cite{Ma:2006km,Barbieri:2006dq,LopezHonorez:2006gr,Press:1985ug}.
The IDM is a model with two Higgs field, one odd under a discrete
symmetry which is introduced to prevent FCNC. The extra Higgs has
also couplings to electroweak gauge bosons but  decoupling of the
extra scalars states (one neutral and one charged in the present
case) one has just to make sure not to violate LEPI bounds on
isospin breaking. This is turns out to be natural in the IDM, thanks
to a hidden custodial $SU(2)$ symmetry.\cite{Hambye:2007vf}.

The  phenomenology of the theory (\ref{lag}) has been much discussed in the literature,
but the focus has somewhat been on heavy or moderately heavy dark matter candidates,
say in the 50 GeV to a few  TeV range. In the present proceeding we consider
a lighter candidate, with $m_S \lsim 10$ GeV. This possibility, first raised in the context of supersymmetric models\cite{Bottino:2003iu,Bottino:2003cz,Bottino:2007qg},  has received
less attention than, say, models with candidates heavier than $50$ GeV, but is nevertheless both viable and phenomenologically
very interesting.

As in the works just quoted, one of our motivation is the  DAMA/NaI and DAMA/Libra experiments,
which have observed an annual modulation in the rate of single
scattering events, with 8.2 $\sigma$ significance (combined)
\cite{Bernabei:2008yi}. This signature is supposed to be one of the
landmark of dark matter-nucleon interactions, the modulation being
due of the combined motion of the Sun and of the Earth with respect
to the halo of dark matter of the Galaxy (generically considered to
be non-rotating) \cite{Freese:1987wu}. The DAMA/NaI detector, and
its successor, DAMA/Libra, consist of sodium iodide (NaI) crystals
and use scintillation to measure the nuclei recoil energy. While the
all the other dark matter experiments work hard on eliminating the
possible background, the strategy of  DAMA is essentially  to
exploit the possible annual modulation. Most of the background is
supposedly eliminated by focusing on single hit events, but of
course contamination by mundane radioactivity is still expected to
exist. The DAMA data are impressive and no explanation but dark
matter really exists. Nevertheless the interpretation in terms of
elastic collisions of nuclei in the detector with dark matter from
the halo is challenged by the null results of various other direct
detection experiments, at least those that are probing similar dark
matter  mass and cross section ranges.

Elastic  scattering has been addressed in various works, possibly
taking into account the possible uncertainties regarding the
properties of the dark matter halo, or of the interaction of dark
matter with nuclei
\cite{Petriello:2008jj,Savage:2008er,Fairbairn:2008gz,Savage:2009mk}.
All these works assert  that the interpretation of the DAMA results
in term of the elastic scattering of dark matter is inconsistent
with the null results of other direct detection experiments (CDMS
\cite{Akerib:2005kh,Ahmed:2008eu}, XENON10 \cite{Angle:2007uj},
TEXONO \cite{Lin:2007ka}, CRESST\cite{Angloher:2002in}, COUPP
\cite{Behnke:2008zza} and CoGenT \cite{Aalseth:2008rx}), at least if
all events the DAMA events are 'quenched'. Quenching relates to the
fact that, after a collision with dark matter, a recoiling nuclei
may loose energy either through electrons (which are responsible for
scintillation) or through collisions with other nuclei (heat,
phonons). The energy measured (which is expressed in electron
equivalent keV, or keVee) is thus typically smaller than the true
nuclei recoil energy, $E_{ee} = Q E_{\mbox{\rm recoil}}$, where $Q$
is the so-called quenching factor,  $Q\lsim 1$.  In a crystal
however, like in NaI, some of the recoiling nuclei events may occur
along the axis of the crystal, in which case  collisions with nuclei
are ineffective and $Q \approx 1$. This effect is called channelling
\cite{Bernabei:2007hw,Bottino:2007qg}. If channelling is taken into account, it may
be possible to explain the DAMA results (at 3$\sigma$) with a light
$m_{DM} \lsim 7-8$ GeV candidate, with a cross section (normalised
to a nucleon) $\sigma_n \approx 10^{-5}$ pb
\cite{Petriello:2008jj,Savage:2009mk}. One should emphasise that the
fit to data is not very good since the chi-square is minimum for
values which are excluded by the other experiments. Furthermore this
interpretation implies that the background is small in the relevant
region of recoil energies. This is embarrassing but, yet, the
possibility that many models, including supersymmetric ones, may
actually explain the DAMA data is not excluded
\cite{Foot:2008nw,Bottino:2008mf,Dudas:2008eq,Feng:2008dz,Andreas:2008xy,Kim:2009ke,Kumar:2009af}.
Although we do not address the possibility here, we should also
mention the very interesting possibility of explaining the DAMA data
(with a substantially improved fit) in terms of the inelastic
scattering of dark matter particles
\cite{Chang:2008gd,TuckerSmith:2001hy,Cui:2009xq,MarchRussell:2008dy}.
This requires some finely tuned models, but beautifully fits the
data.

In the sequel, we focus on elastic scattering and report on the phenomenology
of a singlet scalar (including the IDM)\cite{Andreas:2008xy}.
For the sake of reference, we refer to the range quoted
in the work of Pietrello and Zurek\cite{Petriello:2008jj},
corresponding to
$$
3 \times 10^{-5} \mbox{\rm pb} \lsim \sigma_n^{SI} \lsim 5 \times 10^{-3} \mbox{\rm pb},
$$
and
$$
3 \mbox{\rm GeV} \lsim m_{DM} \lsim 8 \mbox{\rm GeV}.
$$
This range, which is based on the two bins version of the DAMA data
for the spectrum of modulated events \cite{Petriello:2008jj}, is
clearly larger than the one based on the full set of data
\cite{Savage:2009mk}. If anything, like if channelling turns out to
be in-operant, this range is representative of a class of models
which are not yet excluded by any experiments.

\section{Direct Detection and Relic Abundance}

For the model (\ref{lag}), the only processes  relevant for direct
detection and to fix the relic abundance (we consider the standard
thermal freeze-out) are those of Fig.1. This is also true for the
IDM provided $m_{DM} \ll m_{Higgs}, m_{Z,W}$. Since these processes
have the same dependence in the coupling $\lambda_L$ and the Higgs
mass, both processes are closely related
\cite{Burgess:2000yq,Andreas:2008xy}. Concretely, if we fix the
cross section assuming a cosmic abundance given by cosmological
observations (and thermal freeze-out), then the direct detection
cross section is fixed, modulo the uncertainty in the coupling of
the Higgs to nucleons, parameterised by $f$, with $f m_{N}\equiv
\langle {N}| \sum_q m_q \bar{q}q|{N}\rangle=g_{h{NN}} \mbox{\rm v}$.
Here we take $f=0.30$ as central value, and vary it within the
rather wide range $0.14  < f <  0.66$ \cite{Andreas:2008xy}.
\begin{figure}
\begin{center}
\vspace{2mm}
\epsfig{file=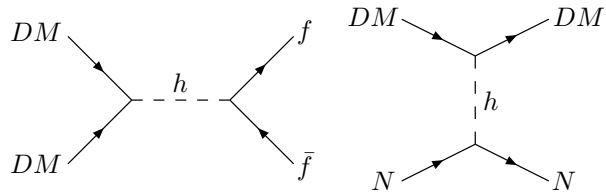,width=8.45cm}
\vspace{-2mm}
\caption{Higgs exchange diagrams for the DM annihilation (a) and scattering with a nucleon (b).}
\vspace{-4mm}
\end{center}
\end{figure}
As Fig.2 shows, compatibility between DAMA (regions in red) and WMAP
(black region) is possible. Conversely, one may say that the red
regions are not excluded by existing dark matter detection
experiments. Notice that the coupling $\vert\lambda_L\vert$ tend to
be large, but is still perturbative. Incidentally, from
(\ref{massS}) we see that $\mu_2 \sim v$ so that a light $m_{DM}$
potentially poses a (small) hierarchy problem. This may perhaps
explain  why this simple model is systematically overlooked in the
current literature even though it has many interesting
consequences.
\begin{figure}
\begin{center}
\epsfig{file=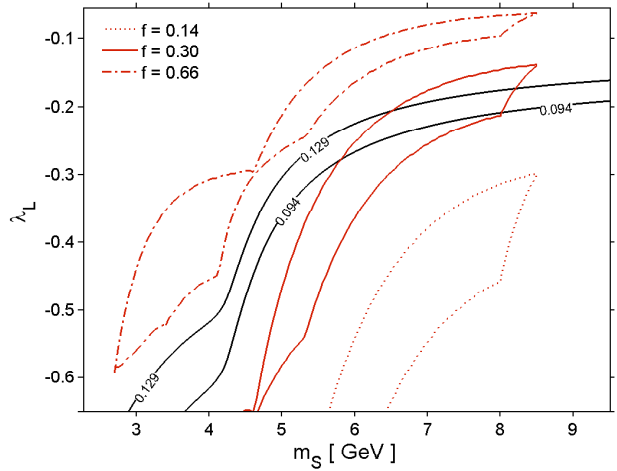,angle=0,width=8.4cm} \caption{For
$m_h=120$~GeV, values of $m_S$ and $\lambda_L$ consistent with WMAP
 $0.094< \Omega_{DM} h^2 < 0.129$ (solid black lines), and which
match the direct detection constraints (two bins fit to DAMA of
Petriello and Zurek \cite{Petriello:2008jj}).}
\end{center}
\end{figure}

\section{Indirect Detection}

The dark matter candidates considered here have both large cross
sections and a large abundance compared to more mundane, heavier
dark matter candidates and their annihilation in the Galaxy may
lead to  some interesting signals.

In Fig. 3 we show the flux from the annihilation of the dark matter
candidate into photons at the centre of the Galaxy.  The mass  of
the DM candidate puts it in the energy range of EGRET data and of
the Fermi/Glast satellite. Fig.3 we show the predicted flux of gamma
rays from the galactic centre for a sample of scalar DM with
parameters which are consistent both with DAMA and WMAP and we
compare to EGRET data \cite{Hunger:1997we,Andreas:2008xy}. It is
interesting that the predicted flux is of the order of magnitude of
the observed flux at the lowest energies that have been probed by
EGRET. A tentative conclusion is that observations by Fermi/GLAST
might constrain the model, modulo the usual uncertainties regarding
the profile of the dark matter at the centre of the Galaxy. Similar
predictions have been reached in the framework of so-called WIMPless
models (see also the talk by Jason Kumar at this conference)
\cite{Feng:2008dz,Kumar:2009af}.

Dark matter may also be captured in the core of the Sun, where its
annihilation may be observed by neutrino detectors. In Fig. 4 we
show the limit which may be expected from Super-Kamiokande
\cite{Andreas:2009hj,Feng:2008qn,Savage:2008er}. The dominant source
of neutrinos is annihilation, through the Higgs, into tau-antitau
pairs (see also the talk by Kumar at this conference)

\begin{figure}
\begin{center}
\epsfig{file=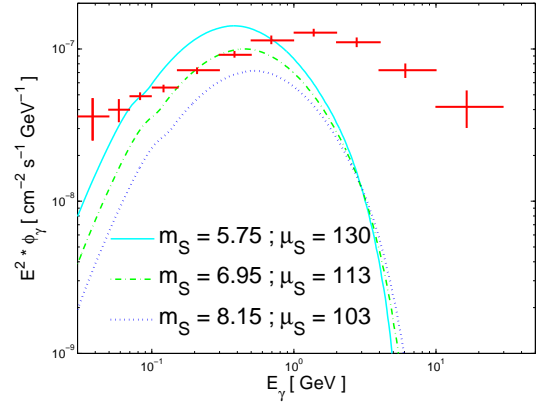,angle=270,width=8cm}
\vspace{-2mm}
\caption{Flux of gamma rays from the galactic center from the annihilation
of a scalar DM consistent with DAMA, compared with EGRET data ($m_h=120$ GeV
and using a NFW profile).
}
\end{center}
\end{figure}

\begin{figure}
\begin{center}
\epsfig{file=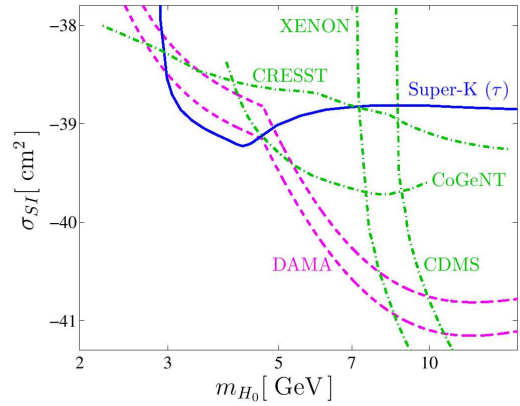,angle=0,width=8cm}
\vspace{0mm}
\caption{Region allowed by DAMA (dashed magenta) together with currents limits on the SI cross section  from direct detection experiments and the limit from Super-Kamiokande (solid blue, mostly from $\bar \tau-\tau$ annihilations).
}
\end{center}
\end{figure}

\begin{figure}
\begin{center}
\epsfig{file=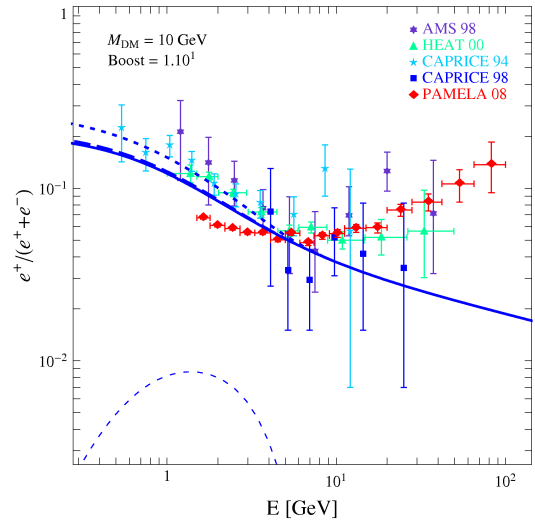,angle=0,width=7cm}
\caption{Positron fraction compared to Pamela and other relevant data.
The signal is weak and within the region affected by solar modulation.
We boost the signal by a factor of 10 (short dash blue line).}
\end{center}
\end{figure}

\begin{figure}[htb!]
\begin{center}
\epsfig{file=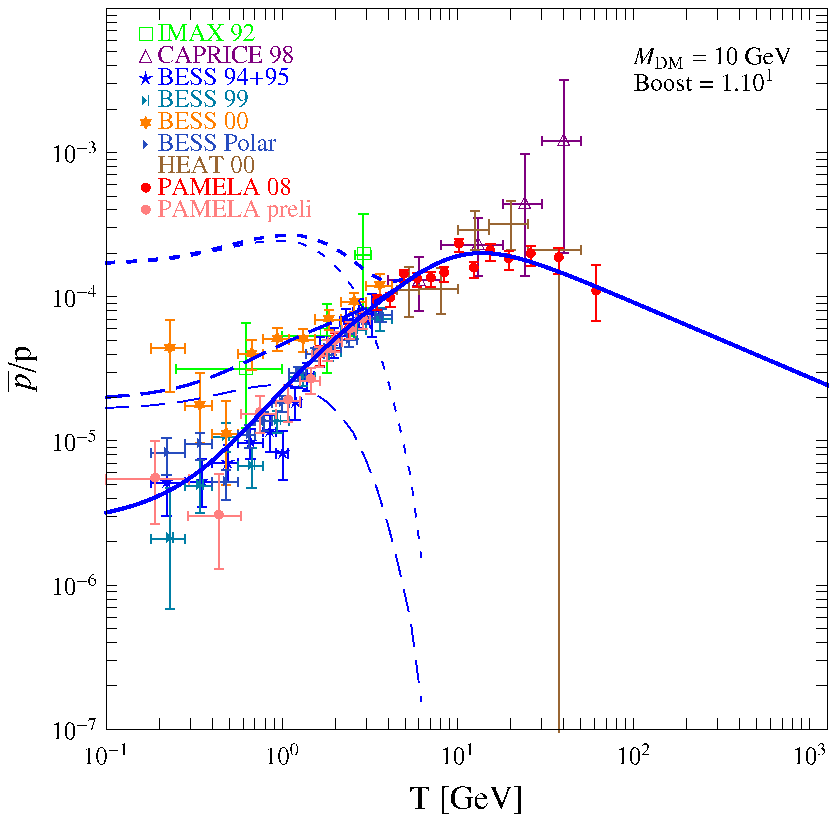,angle=0,width=7cm}
\caption{Flux of antiprotons to protons for a $m_{DM} = 10$ GeV scalar singlet.
We show the contribution for no boost factor (long dashed) and for a boost factor
of 10 (short dashed). The latter is clearly excluded.}
\end{center}
\end{figure}
Finally we may consider the production and propagation of
antiparticles coming from the annihilation of dark matter in the
Galaxy, a possibility which may be constrained using the new data on
the positron and anti-proton fluxes in cosmic rays.  The flux of
positrons and antiprotons is quite large for the candidates
considered here (for the same reasons given at the beginning of this
section the flux of positrons (and for that matter, other cosmic ray
components, including antiprotons). However the fluxes fall in an
energy range where solar modulation severely limits the
possibilities to  constrain the model. In Fig.5 we show the
predictions of the model for the positron fraction, and Fig.6 shows
the $\bar p/p$ ratio \cite{Nezri:2009jd}. The signal in positrons is
rather weak, unless there is a boost factor (say BF=10). The
signature into antiproton is however quite significant and, for
instance, a boost factor of order 10 is clearly excluded, a
conclusion which is probably robust, even without knowing precisely
the impact of solar modulation. A better understanding of solar
modulation is nevertheless desirable  and  perhaps more severe
constraints could be obtained in this way.  Interestingly the model
also predicts a substantial production of antideuteron
\cite{Nezri:2009jd}. The production of anti-deuteron by spallation
in cosmic rays is typically small and is predicted to fall for
kinetic energies below 1 GeV per nucleon. Given its large abundance,
a light WIMP may give a substantial contribution to the flux of
anti-deuteron at low energies \cite{Donato:1999gy,Bottino:2008mf}.
For the IDM candidate with $M_DM$ = 10 GeV, we obtain, using the
DarkSUSY routines, an anti-deuteron flux at $T_{\bar D} = 0.25 $
GeV/n  of $9\cdot 10^{-7}$ (GeV/n s sr $m^2$)) (for BF = 1), which
is below the upper limit of $1.9\cdot 10^{-4}$ (GeV/n s sr $m^2$)
set by the BESS experiment, but above the expected acceptance of the
future AMS-02 and GAPS experiments, which are $4.5 \cdot 10^7$
(GeV/n s sr $m^2$) and $1.5\cdot 10^7$ (GeV/n s sr $m^2$)
respectively. Thus anti-deuteron data might turn out to give the
strongest constraint on the light WIMP dark matter candidate
considered here.

\section{A Light Scalar At The LHC}

In the present model, the coupling between the Higgs and the dark
matter particle is large. This leads to a large Higgs boson decay
rate to a scalar DM pairs at the LHC
\cite{Andreas:2008xy,Burgess:2000yq}.  For example, for $m_S = 7$
GeV and $\lambda_L = -0.2$ and for a Higgs of mass 120 GeV we get
the branching ratio $BR(h \rightarrow SS) = 99.5\%$, while for $m_h
= 200$ GeV and $\lambda_L = -0.55$ we get $BR(h \rightarrow SS)=
70\%$. This reduces the visible branching ratio accordingly,
rendering the Higgs boson basically invisible at LHC for $m_h = 120$
GeV, except possibly for many years of high luminosity data taking.
Such a dominance of the invisible DM channel is a clear prediction
of the framework, although it poses a challenge to experimentalists
\cite{Eboli:2000ze}.

\section{Conclusion}

A light, $m \sim $ few GeV, singlet scalar dark matter candidate interacting
through the Higgs is perhaps not very motivated theoretically speaking
but it may have a very interesting phenomenology.
If its relic abundance is fixed by thermal freeze-out,
all its cross sections are also essentially fixed.
Its elastic scattering with nucleons is in a range consistent
with the modulation observed by DAMA and its annihilation into
various by-products are within reach of current gamma ray, neutrinos
and cosmic ray detectors. A rather clear cut prediction
is the production of a rather large flux of antideuteron in cosmic rays.
Also, it predicts that  a light  Higgs is essentially invisible at the LHC.

\section*{Acknowledgements}

The results reported here are based on work done in collaboration
with Sarah Andreas, Thomas Hambye, Emmanuel Nezri, Quentin Swillens
and Gilles Vertongen. The work of the author is supported by the
FNRS and the Belgian Federal Science Policy (IAP VI/11).

\bibliographystyle{ws-procs9x6}
\bibliography{IDMref}

\end{document}